# Climate-Driven Supply-Demand Matching Reshapes Global SolarEV Decarbonization Potential

Yuchen Liu[1]*, Stefan Pfenninger-Lee[2], Tosiyuki Nakaegawa[3], Takuro Kobashi[1]*
[1] Graduate School of Environmental Studies, Tohoku University, Sendai, Japan
[2] Faculty of Technology, Policy and Management, Delft University of Technology, Delft, Netherlands
[3] Meteorological Research Institute, 1-1, Nagamine, Tsukuba, Ibaraki 305-0052, Japan
* Corresponding authors: liu.yuchen.r2@dc.tohoku.ac.jp; takuro.kobashi.e5@tohoku.ac.jp

## ABSTRACT

Rooftop photovoltaics (PV) can support rapid urban decarbonization, but their contribution is constrained by the imbalance between PV generation and demand, both shaped by local climate. Growing urban electric-vehicle (EV) deployment creates distributed storage near rooftop PV, shifting PV electricity across time. Here we simulate rooftop PV integrated with EVs as batteries across 5,000 grid cells globally, using hourly meteorology and climate-induced demand. We find that PV-EV systems can supply up to 84% of annual building and mobility electricity demand. Self-sufficiency is highest where daytime demand aligns with generation and seasonal variability remains limited. Using PV capacity factor alone to screen PV-EV potential undervalues 21.9% of the world's 1,000 most populous cities because generation-side metrics overlook the temporal coupling of PV generation, electricity demand and EV storage. Substantial PV-EV potential is concentrated in many low-latitude cities, offering a pathway to cleaner urban energy systems while accelerating progress towards carbon neutrality.

**Keywords:** PV-EV systems; rooftop photovoltaics; electric vehicles; vehicle-to-home; self-sufficiency; supply-demand matching

2024 was the first year on record in which global mean temperature exceeded 1.5 °C above pre-industrial levels[1], underscoring the urgency of rapid energy-system decarbonization. Photovoltaics (PV) and electric vehicles (EVs) are expanding rapidly and could play central roles in this transition[2-4]. Yet, their growth also creates a fundamental temporal challenge: PV generation varies with weather and is unavailable after sunset[5-8], while EV charging demand can peak when commuters return home in the evening[9-11]. As their penetration increases, this mismatch between daytime PV generation and non-solar demand can increase the balancing burden on power systems, making their coordinated integration increasingly important.

The SolarEV City concept is built on turning this temporal mismatch into system-level complementarity[12]. Its key lies in using EVs as distributed storage close to rooftop PV. Temporally, EVs can absorb clean electricity from rooftop PV during daylight hours and release it during non-solar periods to meet evening and night-time demand[13,14]; spatially, EVs move between homes and workplaces, potentially carrying stored electricity across locations[10,15]. This flexibility can be enabled through bidirectional Vehicle-to-Home/Vehicle-to-Building (V2H/V2B) operation[16]. In this way, rapidly growing transport electrification need not remain a source of additional electricity demand but can instead become distributed mobile storage that connects daytime PV generation with non-solar demand.

Previous studies have demonstrated substantial system benefits from integrating PV and EVs[9,17-20]. Direct PV charging can substantially improve the environmental performance of EVs, while bidirectional charging can further reduce carbon emissions and energy costs[21-28]. However, these studies typically retain local differences in load structure, EV penetration, charging technologies, rooftop availability and other infrastructure conditions. This heterogeneity obscures how climate, as a fundamental physical driver[29], shapes the matching of PV generation, demand and EV storage, and hinders consistent comparison of PV-EV self-sufficiency potential [30].

Moreover, conventional generation-side indicators, such as rooftop PV capacity and solar-resource availability[31-35], cannot fully characterize the potential of a PV-EV system because they do not capture when PV electricity is generated relative to demand, or how EV storage can shift surplus PV electricity across time. Self-sufficiency provides a system-level metric for PV-EV systems by quantifying the fraction of building and mobility electricity demand met directly by local PV or indirectly through EV time-shifting[36-41], thereby capturing the matching of generation, demand and EV storage.

Here, we use 25-year lifetime-average self-sufficiency rate (hereafter self-sufficiency ratio) as the core metric to quantify climate-induced self-sufficiency potential within a controlled simulation framework. We further examine the consequences of using solar resources alone to screen PV-EV potential. To isolate climatic effects, we use Kyoto as a standardized reference for an urban energy system and hold rooftop area, EV ownership, travel behavior and electricity-use characteristics constant while projecting the same socio-technical configuration onto 5,000 global land-grid locations. Under this controlled experiment, local climate, and the hourly variations in PV generation and demand that it induces, are the only source of spatial variation. Hourly simulation then allows us to capture how PV generation, demand and EV storage synchronize across diurnal and seasonal timescales.

## A standardized reference system for global comparison

We developed a Kyoto-anchored controlled simulation framework to quantify how climate alone shapes PV-EV self-sufficiency ratio globally (Fig. 1). 5,000 global land grid cells were sampled, with local hourly meteorology driving both PV generation and climate-responsive demand. Across all grid cells, the same PV-EV system configuration and dispatch assumptions were retained, allowing differences in self-sufficiency to be attributed to climate-driven differences in the hourly and seasonal matching of PV generation, demand and EV storage.

Kyoto was selected as the reference system for both empirical and physical grounds. The city has served as an established testbed for PV-EV integration, with previous studies demonstrating PV-EV technical, environmental and economic performance from household to urban scales[13,14,42]. Its mid-latitude climate also captures both heating- and cooling-sensitive demand regimes[43], enabling the full range of the climate-demand response to be characterized within a single system[42]. Importantly, Kyoto is not an exceptional electricity system: its normalized thermal neutral weekday profile showed no substantial deviation from the observed range across 11 external systems (Supplementary Fig. 4 and 5). Kyoto therefore provides an empirically grounded, physically informative and externally validated standardized baseline for the global controlled experiment.

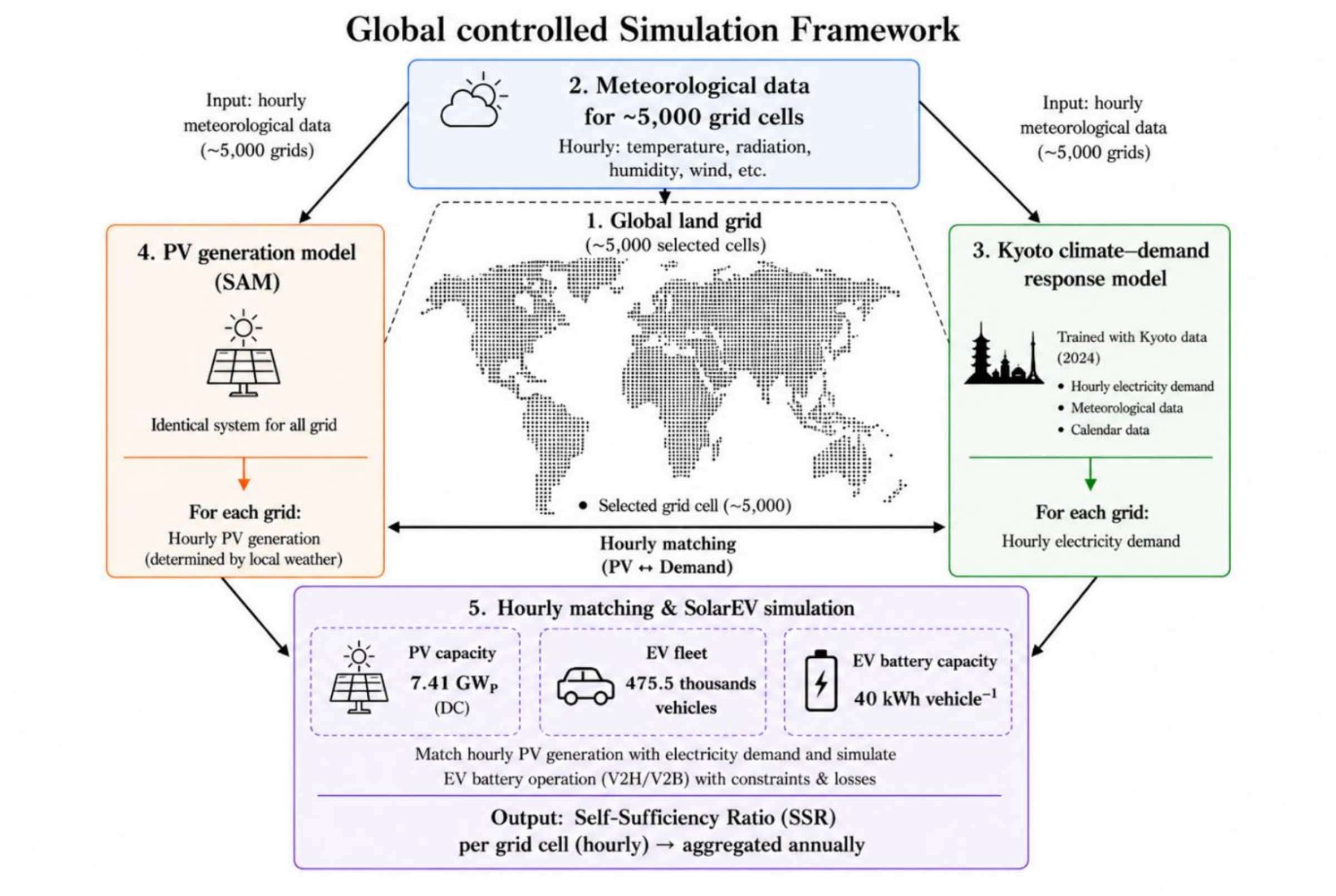

**Fig. 1 | Global controlled simulation framework for PV-EV self-sufficiency.**
5,000 global land grid cells were selected and local meteorology was used to simulate hourly PV generation with the System Advisor Model (SAM) and, through a climate-demand response model trained on Kyoto demand, meteorological and calendar data for 2024, to generate hourly demand. PV generation and demand were then matched hourly under a common PV-EV system configuration, comprising 7.41 GWp of PV capacity, an EV fleet of 475.5 thousand vehicles and a battery capacity of 40 kWh vehicle$^{-1}$. PV-EV operation, including V2H/V2B constraints and losses (Supplementary Information 1.6), was simulated at hourly resolution to derive the self-sufficiency ratio for each grid cell.

**Solar resource alone cannot explain PV-EV self-sufficiency**

Using 8,760 hourly time steps, we simulated the self-sufficiency ratio of the PV-EV system

(Fig. 2a). All grid cells used the same standardized Kyoto-sized configuration: 7.41 GWp of rooftop PV and 475.5 thousand EVs, each with a 40 kWh battery and 6-kW bidirectional V2H/V2B capacity. Self-sufficiency ratio under the PV-EV scenario ranged from 21.6% to 84.2% worldwide and exceeded 60% across most regions south of 45° N (Fig. 2a). High solar resources were concentrated in North Africa, the Middle East and inland Australia, yet high self-sufficiency ratio also occurred along the Mediterranean coast, from the southern United States to northern Mexico, across southern Australia and in several low-latitude coastal regions (Fig. 2a).

For comparison, we also simulated a PV-only configuration. Under the PV-EV scenario, EV batteries provided intraday storage through V2H/V2B operation, shifting surplus daytime PV generation to non-solar and peak-demand periods. The PV-EV system achieved a 21.1-percentage-point higher median self-sufficiency than PV-only (44.8%), despite added EV demand, reaching nearly 80% in some regions (Fig. 2b). However, solar resource alone cannot explain the resulting spatial pattern. In both the PV-only and PV-EV scenarios, regions with similar annual PV productivity exhibited markedly different self-sufficiency ratio, with systematic variation with latitude even within narrow ranges of PV capacity factor (Fig. 2c).

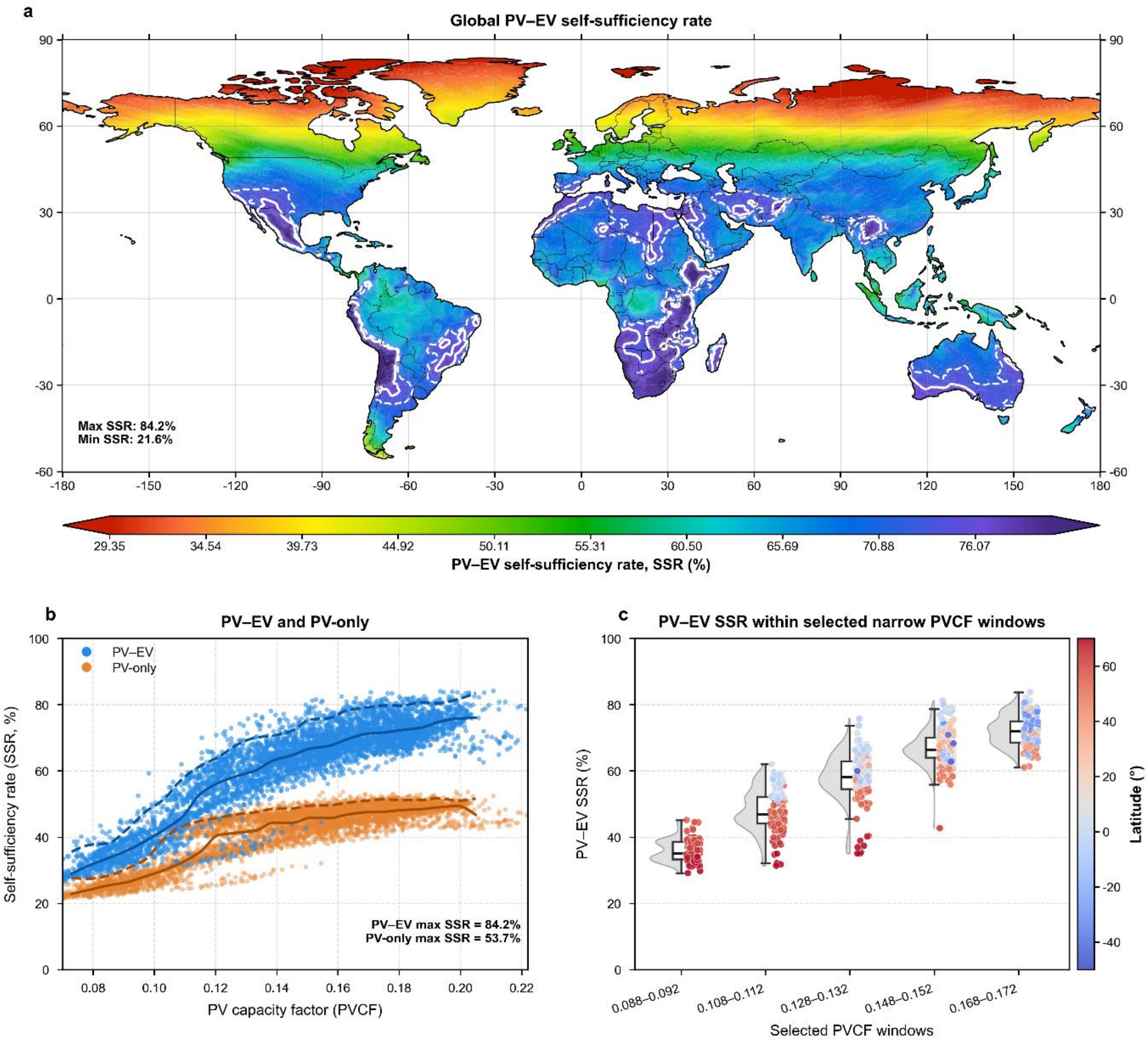

**Fig. 2 | Global PV-EV self-sufficiency and its relationship to PV capacity factor. a,** Global distribution of the PV-EV self-sufficiency ratio across 5,000 land grid cells under the baseline global climate-transfer specification. Colors indicate self-sufficiency ratio, with warmer colors denoting lower values and cooler colors higher values. The dashed white contour delineates the top 20% of grid cells by self-sufficiency ratio. **b,** Relationships between PV capacity factor and self-sufficiency ratio for the PV-EV and PV-only configurations. **c**, Distribution of PV-EV self-sufficiency ratio within five selected narrow PV-capacity-factor windows; points are colored by latitude, showing the remaining latitude-associated variation at similar annual PV productivity.

## Solar resource, demand alignment and demand pressure jointly determine PV-EV self-sufficiency

To explain the spatial variation in PV-EV self-sufficiency ratio, we identified three first-order controls. Solar resource (PV capacity factor), demand alignment (Daytime load), and demand pressure (annual demand) together explained 98% of the spatial variation (Supplementary Note 4.8). These controls represent, respectively, the availability of PV electricity, its temporal alignment with demand, and the pressure placed on available EV storage. Median-based classification of these three variables yielded eight system archetypes, each representing a distinct combination of these controls (Fig. 3a-h)[44].

Solar resource sets the broad boundary of self-sufficiency but does not determine it alone. Near-ideal and Compound-constrained regions accounted for 20.8% and 23.7% of global grids, with median self-sufficiency ratio of 75.1% and 38.6%, respectively (Fig. 3a,h). In resource-abundant regions, high demand pressure or poor demand alignment reduced self-sufficiency ratio by 5.9 and 6.1 percentage points; together, they produced an 11.0-point reduction. Conversely, favorable demand alignment maintained self-sufficiency ratio of 67.7% even in resource-limited regions, comparable to 68.9% observed in resource-rich but poorly aligned systems. Demand pressure had its largest effect when both solar resource and demand alignment were poor, where its self-sufficiency penalty reached 17.7-percentage-point, compared with 4.0-5.9 percentage points across the other archetype contrasts. Thus, solar resource sets the resource ceiling, demand alignment governs how effectively PV generation is used, and demand pressure amplifies the effects of resource and timing constraints.

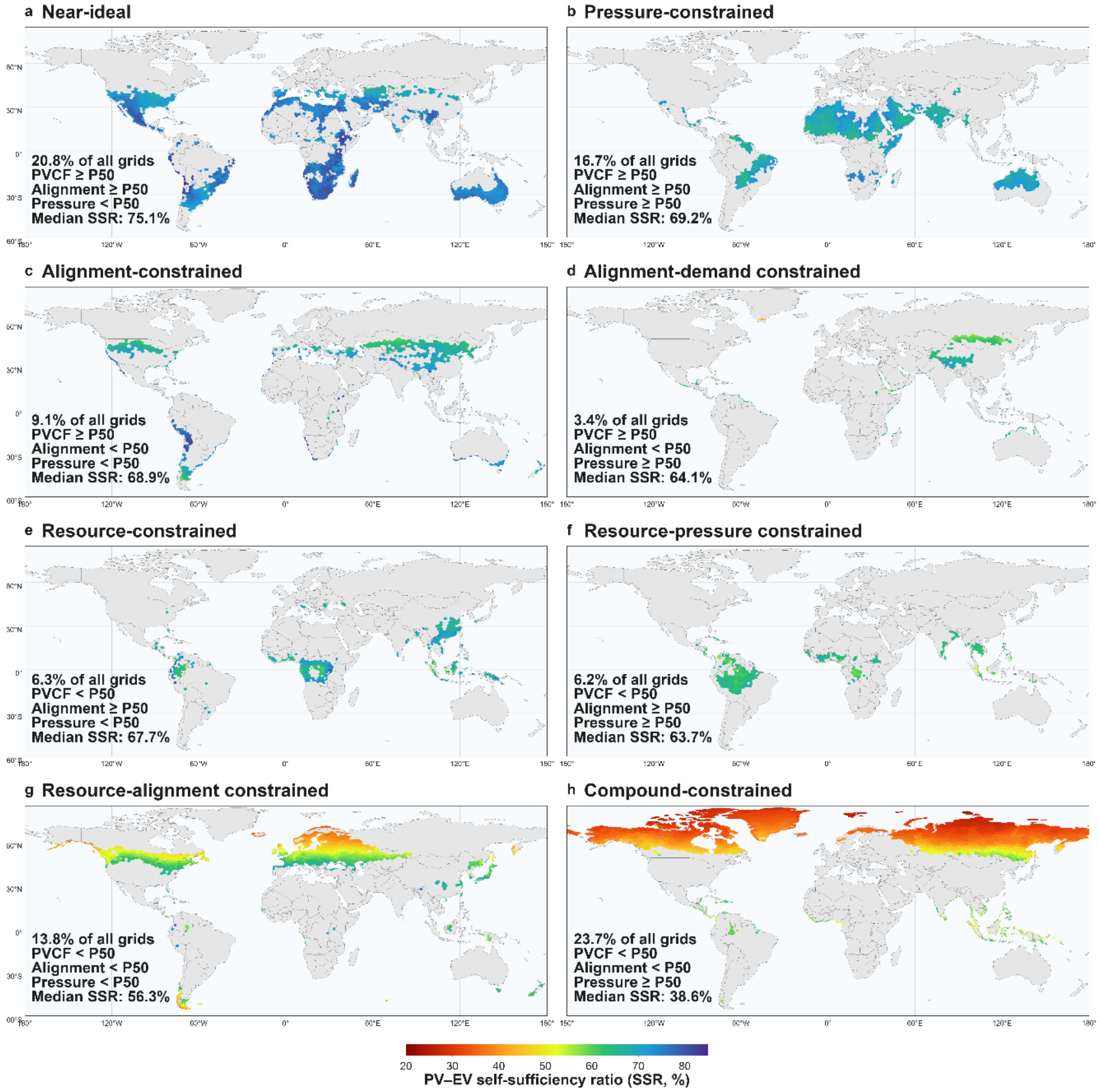

**Fig. 3 | Eight PV-EV operating archetypes defined by PV capacity factor, demand alignment and demand pressure.**

**a-h**, Global distribution of grid cells assigned to eight mutually exclusive operating archetypes. The archetypes are defined by whether PV capacity factor, demand alignment and demand pressure fall above or below their respective global medians. **a**, Near-ideal: high solar resource, high demand alignment and low demand pressure. **b**, Pressure-constrained: high solar resource, high demand alignment and high demand pressure. **c**, Alignment-constrained: high solar resource, low demand alignment and low demand pressure. **d**, Alignment-demand constrained: high solar resource, low demand alignment and high demand pressure. **e**, Resource-constrained: low solar resource, high demand alignment and low demand pressure. **f**, Resource-pressure constrained: low solar resource, high demand alignment and high demand pressure. **g**,

Resource-alignment constrained: low solar resource, low demand alignment and low demand pressure. **h**, Compound-constrained: low solar resource, low demand alignment and high demand pressure. Colors indicate PV-EV self-sufficiency ratio within each archetype; the share of all grid cells and the median self-sufficiency ratio are annotated in each panel.

## Climate controls PV-EV self-sufficiency ratio

Fig. 3 shows that the regions under different latitudinal and climatic conditions tend to fall into distinct PV-EV operating archetypes. The Köppen climate classification gives an important basis for interpreting the climatic mechanisms underlying these archetypes.

Here, we focus on the contrasting Near-ideal and Compound-constrained archetypes, using eight cities across different latitudes and climates as geographic reference points rather than representations of actual socio-economic or infrastructure conditions (Fig. 4). These cases illustrate how climate shapes supply-demand matching and thereby PV-EV self-sufficiency.

Beyond differences in solar resource, demand pressure further distinguishes the two archetypes. Compound-constrained cities carry higher demand, particularly outside solar hours, driven mainly by extreme cooling or heating needs. In Novosibirsk, demand exceeds PV generation even around midday, leaving substantial grid dependence throughout the day (Fig. 4). Port Harcourt, Singapore and Surabaya face a milder but related constraint. Their low latitude provides relatively stable daylight through the year, but monsoon rains in Port Harcourt, daily convective storms in Singapore, and the savanna wet season in Surabaya reduce solar radiation during periods when PV generation could otherwise align with daytime demand (Fig. 4).

Near-ideal cities show the opposite pattern. They combine low demand pressure with favorable alignment between daytime demand and PV generation. Frequent clear-sky conditions support high PV generation in Cairo and Mexico City, while afternoon demand allows daytime surplus to be stored and discharged overnight, sometimes into the following morning (Fig. 4). São Paulo is humid all year, yet its elevation supports relatively favorable solar availability despite its humid climate (Fig. 4). Perth represents another pathway: frontal winter rainfall occurs when solar potential is already low, leaving the high-output summer months relatively unaffected (Figs. 4 and 5).

A further case is the Resource-alignment-constrained class, which is common across maritime Europe and Japan (Fig. 3). Oceanic moderation limits heating and cooling demand, but recurring cloud cover and frontal rainfall reduce solar availability, while mild temperatures weaken daytime demand alignment with PV generation and shift more demand to non-solar hours.

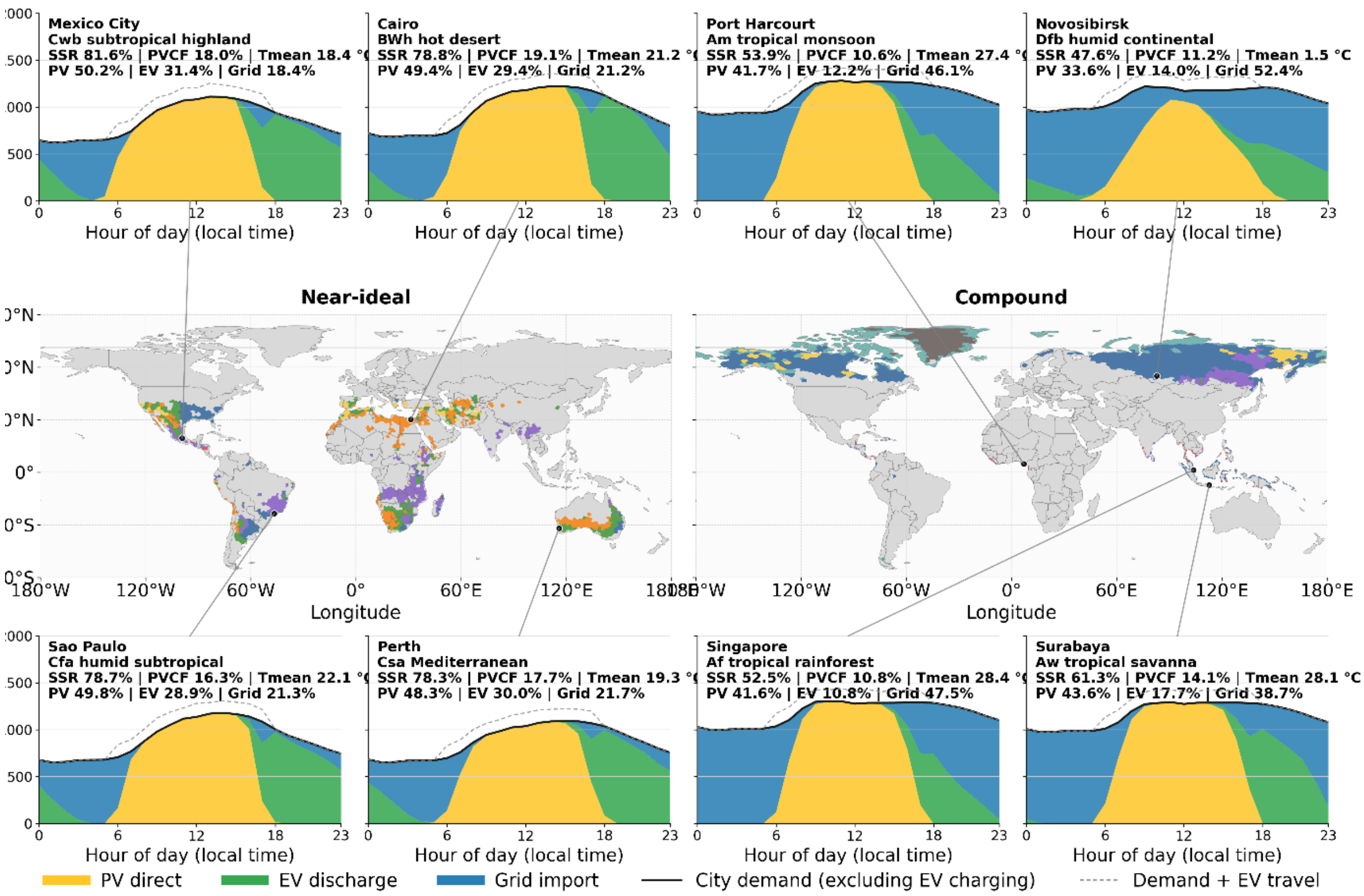

**Fig. 4 | Climatic mechanisms behind the Near-ideal and Compound-constrained archetypes.**

Maps show the global distribution of the Near-ideal (left) and Compound-constrained (right) archetypes, with grid cells colored by the precipitation-related subclass of the Köppen climate classification to highlight hydroclimatic differences beyond the broad thermal gradient associated with latitude. Annotated points mark eight representative cities used to illustrate the underlying mechanisms. City-level values are obtained by spatial interpolation from surrounding grid cells. Mexico City, Cairo, São Paulo and Perth represent the Near-ideal archetype, and Port Harcourt, Novosibirsk, Singapore and Surabaya represent the Compound-constrained archetype. Panels show the mean daily PV-EV energy balance for each city, decomposed into PV electricity consumed directly (yellow), electricity supplied by EV discharge (green), and grid imports (blue), plotted against city demand excluding EV charging load (black line). Each panel reports the PV-EV self-sufficiency ratio (SSR), PV capacity factor (PVCF), mean temperature (Tmean), and the share of demand met by PV, EV discharge and grid import.

## Seasonality controls the stability of PV-EV self-sufficiency

We next examine how seasonality reshapes PV-EV performance across the eight cities (Fig. 5). The mid-latitude arid corridor shows particularly stable supply-demand matching because seasonal changes in PV generation and demand remain favorably aligned. Mexico City and Cairo show this pattern most clearly, maintaining self-sufficiency ratio near 80% all year with little variation and low demand-alignment variability (CValignment = 0.090 and 0.147) (Fig. 5). In the hottest months, their daily demand curve develops pronounced midday peaks, while morning and evening demand change little between the best and worst months. Because the additional cooling demand occurs within the solar window, demand rises when PV output is also high, preserving strong supply-demand matching.

Seasonality weakens the stability of PV-EV self-sufficiency ratio when PV generation and demand vary in opposite directions across seasons, as Novosibirsk illustrates at high latitude. PV surplus appears in summer, when solar elevation angles are highest, but PV output drops sharply in winter, just as heating demand rises and shifts strongly into non-solar hours. Self-sufficiency ratio accordingly drops to about 20-30% in winter before rising to about 90% in summer (Fig. 5). EV batteries cannot compensate for this seasonal mismatch, because the shortfall persists for months rather than hours.

At low latitudes, seasonal variability in PV generation is driven largely by rainfall and cloudiness, while demand timing remains comparatively stable, as Singapore, Port Harcourt and Surabaya illustrate. In Singapore, convective rainfall follows a consistent diurnal pattern throughout the year, and its equatorial location keeps humidity and temperature relatively stable year-round, so demand alignment barely changes and its CV is close to zero (Fig. 5). In Port Harcourt and Surabaya, self-sufficiency ratio drops during their respective wet seasons and recovers during drier months as PV generation increases (Fig. 5).

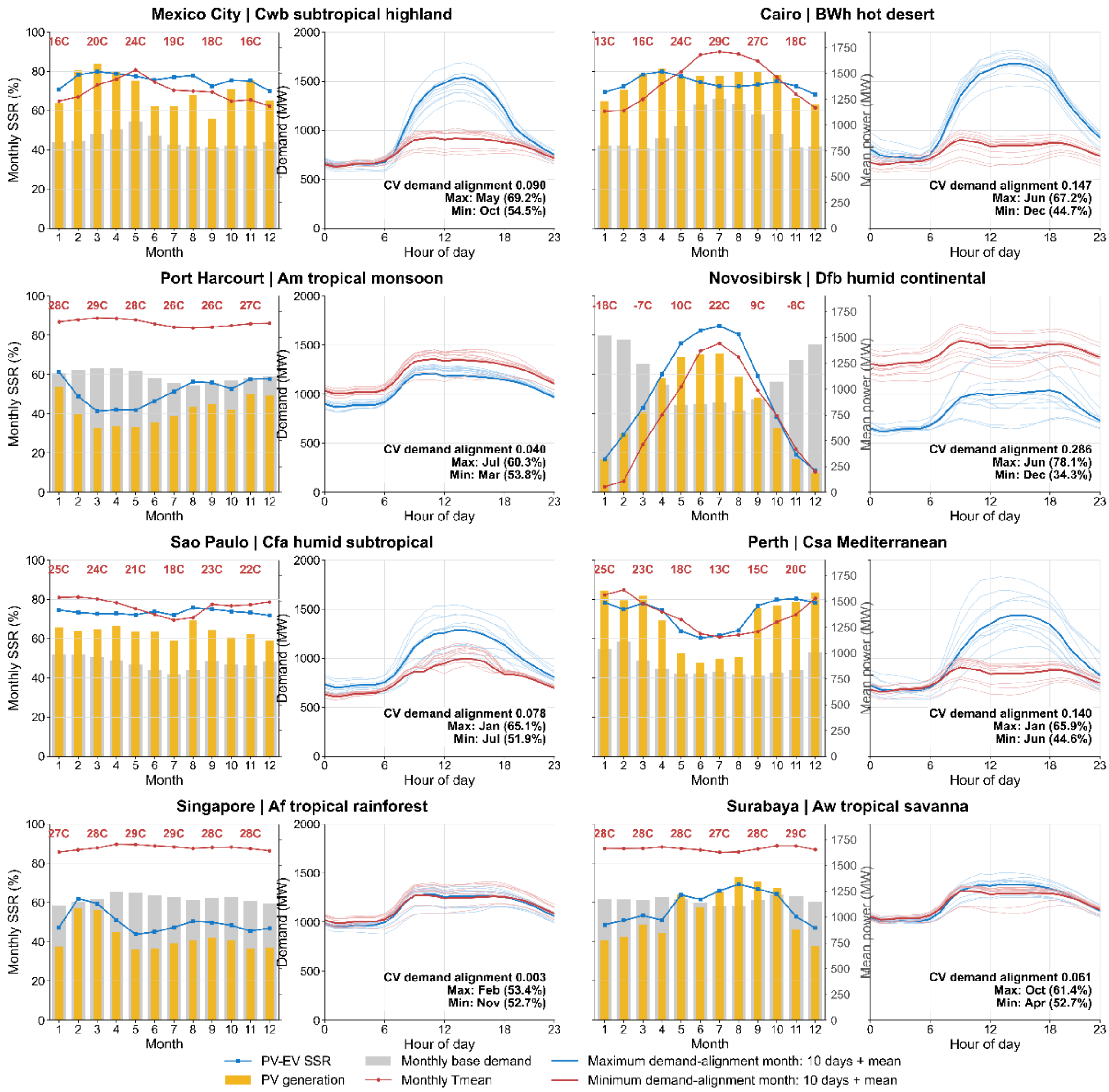

**Fig. 5 | Seasonal stability of PV-EV performance across eight representative cities.**
**a-h**, Monthly PV-EV self-sufficiency ratio (blue line), monthly PV generation (yellow bars), monthly building demand excluding EV charging (grey bars) and monthly mean temperature (red line and labels) for Mexico City, Cairo, Port Harcourt, Novosibirsk, São Paulo, Perth, Singapore and Surabaya (Monthly figure with EV demand shown in Supplementary Note 5.3). Insets show the mean daily profile for the month with the highest demand alignment (blue) and the month with the lowest alignment (red), together with the coefficient of variation of monthly demand alignment and the months of maximum and minimum demand alignment. Low CV values indicate that demand alignment remains seasonally stable, whereas high CV values indicate strong seasonal variability in the timing of demand relative to daytime hours.

**Solar-resource screening overlooks PV-EV self-sufficiency potential in rapidly growing regions**

Screening candidate sites based solely on solar resource (PV capacity factor) can systematically misrepresent PV-EV potential. For each grid cell, we compared the percentile rank of PV-EV self-sufficiency against the percentile rank of PV capacity factor under an identical installed PV capacity. Using a threshold of 15 percentile points to define severe misclassification, PV capacity factor-only screening under-ranks 12.4% of locations and over-ranks 12.4% across the 5,000 global grid cells (Fig. 6). This apparent balance, however, breaks down at the urban scale. Among the world's 1,000 most populous cities, 219 cities (21.9%) are severely under-ranked for PV-EV potential, whereas only 37 cities (3.7%) are over-ranked (Fig. 6).

These misclassified cities exhibit clear regional clustering (Fig. 6b). Of the 219 severely under-ranked cities, East Asia (82 cities), Africa (51 cities), and South America (39 cities) together account for more than three quarters of the total, revealing a pronounced regional concentration of major urban locations that PV capacity factor-based screening systematically under-ranks. Although these regions span diverse climatic settings, most severely under-ranked cities fall within the Near-ideal and Resource-constrained archetypes, where favorable climate-driven demand profiles either amplify the benefits of abundant solar resources or compensate for limited ones. In South America, strong solar resources and favorable generation-demand alignment produce higher PV-EV self-sufficiency than solar-resource screening alone would suggest. Along the Mediterranean, hot, dry summers strengthen this seasonal alignment, whereas in East Asia, Southeast Asia and Central Africa, favorable demand profiles partly offset cloud- and rain-induced reductions in PV capacity factor. Overall, PV capacity factor-based screening severely under-ranks cities when it overlooks the climate-driven matching between PV generation and demand and the seasonal stability.

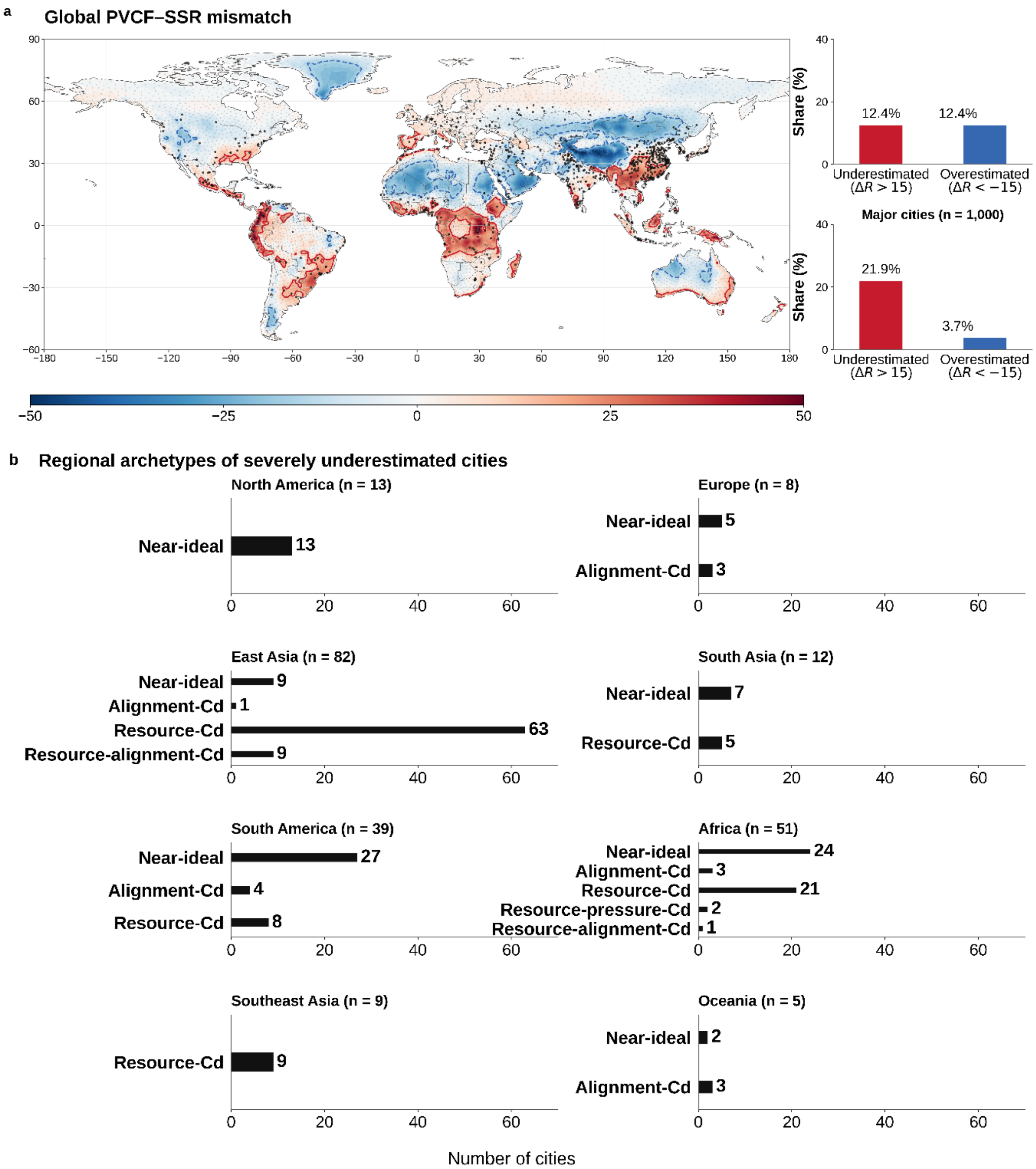

**Fig. 6 | Where PV capacity factor screening misrepresents PV-EV potential.**

**a,** Global mismatch between PV-EV self-sufficiency ratio and PV capacity factor screening across 5,000 global land grid cells. Mismatch is defined as ΔR = percentile rank of PV-EV self-sufficiency ratio − percentile rank of PV capacity factor. Positive values (red) indicate locations where solar resource-only screening under-ranks PV-EV self-sufficiency potential, whereas negative values (blue) indicate over-ranking. Red and blue contours denote severe mismatch (ΔR > 15 and ΔR < −15 percentile points, respectively). Black points indicate the world's 1,000 most populous cities. Insets compare the shares of severely under-ranked and over-ranked grid cells and major cities. **b,** PV-EV operating-archetype composition of the 219 severely under-ranked major cities (ΔR > 15), grouped by world region. Bars show the number of cities in each represented archetype on a common horizontal scale. Archetypes are defined in Fig. 3.

## Discussion

Solar resource alone is insufficient to evaluate PV-EV potential. Using a global controlled simulation that holds social behavior and technical configurations constant across regions, we isolate the climatic determinants of PV-EV self-sufficiency. Using a 15-percentile-point threshold for severe mismatch, PV-resource screening misclassifies roughly one-quarter of global land-grid cells and severely under-ranks the PV-EV potential of 21.9% of the world's 1,000 most populous cities.

In our setup, EVs use V2H/V2B to shift daytime PV surpluses to non-solar periods. In representative near-ideal and Compound-constrained cities, EV discharge meets 10.8-31.4% of total daily demand even with a 40 kWh battery and 12.2-43.9% with an 80 kWh battery (Supplementary Fig. 8). These results highlight the importance of EV temporal flexibility in extending the usable hours of rooftop PV electricity. Generation-side metrics are appropriate for evaluating rooftop PV alone, but coupled PV-EV performance depends on how generation, demand and storage interact over time. Our analysis extends previous approaches by using the self-sufficiency ratio to integrate PV generation, demand alignment and demand pressure into a single outcome of local energy provision, thereby capturing their matching at the system level.[45]

Previous studies have commonly treated seasonality primarily as generation-side variability, driven by precipitation, solar elevation and changes in day length[10,18,46]. Our results show that the system consequences of seasonality depend on whether seasonal variations in PV generation and demand reinforce or offset one another. Regions with comparable annual solar resources can therefore exhibit markedly different PV-EV self-sufficiency when hot, dry summers contrast with climates in which hot and rainy seasons coincide. Seasonal matching performance therefore determines whether climatic variability strengthens or weakens PV-EV performance, extending the conventional generation-centered interpretation of seasonality. More importantly, it determines whether siting decisions based on solar-resource maps become misleading as resource-only indicators omit both the seasonality of demand and its alignment with generation.

This supply-demand matching is unlikely to remain geographically fixed under climate change. Changes in atmospheric circulation and the hydrological cycle may enhance solar availability in some dry regions while reducing it in wetter or monsoon-dominated regions[47]. In East Asia, shifts in the seasonal march may further alter the timing of solar availability relative to electricity demand[48]. Meanwhile, warming may increase cooling demand while reducing heating demand, thereby changing the seasonal and diurnal alignment between PV generation and demand[48]. The archetypes identified here should therefore be interpreted as climate-conditioned system states rather than fixed geographic classes.

Kyoto serves here as a standardized reference case for global comparison rather than as a template for real-world deployment. The baseline configuration corresponds to a PV-to-EV capacity ratio of approximately 0.39 kWp $kWh^{-1}$ and defines the socio-technical conditions represented by the main map. Sensitivity analyses under alternative PV and EV configurations, EV travel patterns and climate-demand responses show that the principal conclusions are robust (Supplementary Note 4).

Under the baseline configuration, an 80% self-sufficiency ratio in the Near-ideal archetype would correspond to approximately 6.81 TWh of local building and mobility demand being supplied by local PV electricity. In high-latitude regions, monthly energy-sufficiency can reach 136.8% in summer (Supplementary Fig. 25), creating potential value for long-duration or seasonal storage. Conversely, prolonged periods of low solar availability can deplete short-duration EV storage, exposing a resilience dimension not captured by the self-sufficiency ratio. This may become particularly important where extreme precipitation or tropical cyclones coincide with disruptions to centralized electricity supply.[49,50]

A central implication is therefore that PV-EV siting should move beyond solar-resource maps towards assessments based on supply-demand matching. This is particularly important for rapidly growing urban regions. Of the 219 cities that are severely under-ranked, more than half are in Africa, South America, South Asia and Southeast Asia; including East Asia raises this share to 88.1%. These regions are expected to absorb a large proportion of future urban population and electricity-demand growth[51,52,53], while basic electricity access remains far from universal. In sub-Saharan Africa alone, around 600 million people still lack access to electricity[54]. At the same time, in rapidly growing regions such as Southeast Asia, much of the electricity demand continues to be supplied by coal- and gas-based systems[55]. Coupled rooftop PV-EV systems may therefore offer particularly favorable climatic conditions for expanding clean urban electricity supply in rapidly growing regions, while providing a common metric for assessing progress towards carbon-neutral urban energy systems.

This study also has clear boundaries. The controlled design isolates climate-driven mechanisms, whereas real-world deployment depends on local load profiles, travel and parking behavior, building stock, distribution-grid capacity, electricity prices and market rules. Our analysis also represents a contemporary climatic baseline, while future climate change may reshape the magnitude and geography of PV-EV self-sufficiency by altering solar radiation, cloudiness and temperature-sensitive demand.[46-48] We focus primarily on the distributed storage provided by EVs, while also considering stationary-battery scenarios (Supplementary Note 5.4). EV availability depends on travel schedules, state of charge, trip demand and user behavior, so available storage cannot be inferred from nominal battery capacity alone. Future work should test whether the identified archetypes persist under observed load and mobility data, future climate conditions and prolonged low-renewable-output events. Such Dunkelflaute conditions may become more consequential in highly renewable systems, with recent studies indicating increasing risk in Japan and comparatively limited change in Germany.[56-57] Further work could also build on this comparison by examining how declining stationary-battery costs reshape the relative value of mobile and stationary storage, and by extending the framework to distribution-grid constraints, V2G market design and broader carbon-neutral urban energy systems, including PV-heat pump-thermal storage and wind-to-hydrogen networks.

Overall, this study shifts the evaluation of PV-EV systems from solar-resource ranking towards a climate-driven supply-demand matching perspective.

## Methods

### Data

This study builds a controlled global PV-EV simulation framework using meteorological data, electricity demand data, spatial grids, rooftop PV potential, vehicle ownership and socio-technological parameters. Meteorological and radiation inputs were obtained from NASA GES DISC MERRA-2 hourly products through OPeNDAP. We used M2T1NXSLV.5.12.4 for near-surface meteorological variables, including 2 m air temperature (T2M), 2 m specific humidity (QV2M), 10 m wind components (U10M and V10M) and surface pressure (PS), and M2T1NXRAD.5.12.4 for surface downward shortwave radiation (SWGDN). The data were extracted for each grid cell, converted from UTC to local time, and used for BAIT calculation (Supplementary Information 2.2), demand reconstruction, demand transfer and PV generation simulation.

The Kyoto demand archetype is built using 2024 hourly load data from the Kansai Electric Power service area, hourly municipal weather data, and annual municipal electricity demand. The hourly regional load data come from the historical regional supply-demand records published by the Organization for Cross-regional Coordination of Transmission Operators. Annual municipal electricity demand comes from the municipality-level electricity demand tables in the Electricity Survey Statistics published by the Agency for Natural Resources and Energy under the Ministry of Economy, Trade and Industry.

The global simulation units are constructed in Python as 5,000 land grid cells. Ocean and non-land areas are excluded. Each land grid cell is treated as an independent simulation unit. Each grid cell is assigned its corresponding MERRA-2 weather series, transferred hourly demand, PV generation profile, standardized rooftop PV potential, and EV ownership assumption.

Hourly PV generation is simulated using the NREL System Advisor Model, or SAM. SAM generates hourly generation profiles for a standard 10 kW PV system under the weather conditions of each grid cell. The PV performance model follows the NREL technical documentation.

Standardized rooftop PV potential and EV ownership assumptions follow the Japanese municipal PV-EV data and assumptions from Trang et al[39]. Rooftop area comes from the official building footprint data provided by the Geospatial Information Authority of Japan. Maximum rooftop PV capacity is calculated using an area-to-capacity conversion factor of $7.0\ \mathrm{m^2\ kW^{-1}}$. Municipal private passenger vehicle ownership comes from vehicle registration data from the Automobile Inspection and Registration Information Association and the Japan Light Motor Vehicle and Motorcycle Association. These data are published by the Ministry of the Environment of Japan in Current Estimates of $CO_2$ Emissions by Sector.

The techno-economic assumptions include PV degradation rate, EV battery capacity, SOC constraints, round-trip efficiency, and EV travel parameters. These parameters are taken from previous SolarEV studies, NREL, IEA/IRENA, Nichicon, Geotab, BloombergNEF, GlobalPetrolPrices, and the technological parameter table compiled in this study. Full values and sources are reported in Supplementary Table 1.

In the following workflow, MERRA-2 meteorology first enters BAIT calculation and SAM-based PV simulation. The Kyoto demand archetype is then reconstructed and transferred to each global grid cell using local meteorology. Finally, the transferred demand and PV generation profiles are combined in hourly PV-EV dispatch under standardized PV-capacity settings.

**Building-adjusted indoor temperature**

We use the building-adjusted indoor temperature index, or BAIT, to represent the combined thermal effect of local weather on electricity demand. BAIT is a daily building-level proxy for perceived indoor temperature. It estimates the thermal environment inside buildings without active heating or cooling. Unlike outdoor air temperature, BAIT considers solar heat gain, wind-driven cooling, humidity-related discomfort, and short-term building thermal inertia.

For each target location, we extract hourly MERRA-2 variables. These variables include 2 m air temperature, specific humidity, 10 m zonal and meridional wind speeds, surface pressure, and surface downward shortwave radiation. Wind speed is calculated from the two horizontal wind components. Because BAIT is defined at the daily scale, all timestamps are first converted from UTC to local time. Daily aggregation is then performed after this conversion. This step ensures that daily temperature, radiation, wind, and humidity statistics are based on the local meteorological day rather than a single reference time.

Daily BAIT is calculated following the Renewable.ninja method. The index adjusts daily mean outdoor air temperature using solar radiation, wind speed, and humidity terms. It then applies short-term thermal-memory smoothing to represent the delayed thermal response of buildings. Under hot conditions, part of the unsmoothed outdoor temperature is mixed back into the smoothed BAIT. This step represents the weakening of building thermal memory caused by ventilation and heat release. Finally, each daily BAIT value is assigned to local noon and interpolated to hourly resolution using cubic splines. The interpolated hourly BAIT is used as a meteorological predictor in the Kyoto hourly-profile reconstruction model and the global climate-transfer model.[58]

**Reconstruction of hourly electricity demand in Kyoto**

Official hourly electricity demand data in Japan are usually available only at the power-region scale. They are not available at the municipal scale. We therefore use a top-down downscaling model to reconstruct hourly electricity demand for Kyoto City from the hourly load curve of the Kansai Electric Power service area. This step does not aim to recover unobserved true municipal hourly electricity use. It aims to generate a Kyoto-style hourly demand archetype that responds to local weather and matches the annual municipal electricity demand total.

The supervised learning target is the observed hourly load shape of the Kansai Electric Power service area in 2024. To avoid directly transferring the absolute regional load scale to municipalities, we first normalize the hourly Kansai load by its annual maximum. During training, regional weather inputs are calculated as the annual-electricity-demand-weighted average of hourly weather series across Kansai municipalities. The meteorological predictors include air temperature, humidity, wind speed, solar radiation, and BAIT.

The model features include hour, month, weekday, weekend indicator, Japan-specific holiday indicator, monthly cyclic terms, and meteorological variables. Temperature-sensitive electricity demand is represented by heating degree hours and cooling degree hours:

$$\mathrm{HDD}_t = \max(0, T_H - \mathrm{BAIT}_t),$$
$$\mathrm{CDD}_t = \max(0, \mathrm{BAIT}_t - T_C),$$

where $T_H$ and $T_C$ are the heating and cooling thresholds. Candidate heating thresholds range from $11°\mathrm{C}$ to $17°\mathrm{C}$. Candidate cooling thresholds range from $17°\mathrm{C}$ to $23°\mathrm{C}$. Threshold pairs with $T_C - T_H < 2°\mathrm{C}$ are excluded. The final threshold pair is selected based on the fitting performance of XGBoost on the normalized Kansai load training set.

The normalized Kansai load shape is reconstructed using XGBoost as the primary model, with LightGBM used as a model-robustness check. XGBoost hyperparameters are optimized using Optuna.

After training, the model is applied to the local hourly weather series of Kyoto and other Kansai municipalities. This step generates dimensionless hourly load curves.

Because tree-based and ensemble models may smooth peak demand, we apply variance correction to the predicted curves. Let $\sigma_{\mathrm{obs}}$ be the standard deviation of the observed normalized Kansai load. Let $\sigma_{\mathrm{pred}}$ be the standard deviation of the predicted normalized Kansai load. The raw variance amplification factor is:

$$f_{\mathrm{raw}} = \frac{\sigma_{\mathrm{obs}}}{\sigma_{\mathrm{pred}}}.$$

The final Kyoto hourly demand curve is used as the reference behavioral profile in the global climate-transfer model. Further details of the Kyoto hourly-profile reconstruction and annual scaling procedure are provided in Supplementary Information 2.1.

**Physically constrained global climate-transfer model**

Two model components are used. The Kyoto hourly-profile reconstruction model recovers the Kyoto reference demand profile from Kansai load and local meteorological and behavioral predictors. The global climate-transfer model then projects this Kyoto-style system under alternative climates while holding its behavioral structure and reference scale fixed. The baseline global-transfer specification used for the main simulations combines a winter BAIT heating term, a summer outdoor-temperature cooling term and a calendar-only XGBoost residual. Two alternative thermal-response specifications, BAIT-only and Temperature-only, are evaluated as sensitivity cases. This transfer process is not used to estimate the real electricity consumption of each country or region. Instead, it evaluates how a fixed Kyoto-style demand structure responds under different global climate conditions.

The global climate-transfer model separates temperature-driven demand from residual time effects. Heating and cooling degree-hour variables are defined as:

$$H_{g,t} = \max(0, T_H - T_{g,t}^{\mathrm{BAIT}}),$$

$$C_{g,t} = \max(0, T_{g,t}^{\mathrm{out}} - T_C),$$

where $T_{g,t}^{\mathrm{BAIT}}$ is the building-adjusted temperature proxy for grid cell $g$ at hour $t$, and $T_{g,t}^{\mathrm{out}}$ is outdoor air temperature. The heating term uses BAIT because heating demand is affected by building thermal inertia and delayed heat transfer through the building envelope. The cooling term uses outdoor air temperature to retain the immediate trigger effect of high temperature on cooling demand.

The model uses a two-stage hybrid structure. In the first stage, an L2-regularized Ridge model estimates the temperature-driven physical baseline demand:

$$D_t^{\mathrm{temp}} = f_{\mathrm{Ridge}}(H_t, C_t).$$

In the second stage, XGBoost estimates the non-temperature residual demand:

$$\epsilon_t = D_t^{\mathrm{Kyoto}} - D_t^{\mathrm{temp}},$$

$$\hat{\epsilon}_t = f_{\mathrm{XGB}}(z_t),$$

where $z_t$ includes hour of day, day of week, weekend indicator. The final transferred hourly demand for grid cell $g$ is:

$$D_{g,t} = \max\left[0, f_{\mathrm{Ridge}}(H_{g,t}, C_{g,t}) + f_{\mathrm{XGB}}(z_{g,t})\right].$$

Validation of the baseline global climate-transfer specification used a chronological split: the first eleven months were used for fitting, and 1-28 December was retained as the normal-operation hold-out period; 29-31 December were excluded because of atypical year-end demand. This restricted specification combines winter BAIT, summer temperature and calendar-only residual predictors. It achieved a hold-out MAPE of 4.37% and $R^2$ of 0.86. These metrics evaluate the baseline global climate-transfer specification; validation of the Kyoto hourly-profile reconstruction model is reported separately in Supplementary Fig. 3.$R^2$

**Hourly global PV-EV simulation**

After obtaining transferred hourly electricity demand for each grid cell, we develop a grid-cell-level hourly simulation model. The model quantifies the self-sufficiency ratio after distributed PV is coupled with EV-based bidirectional storage. Each global land grid cell is simulated independently at a 1-hour resolution. Leap days are excluded, so each simulation year contains 8,760 hours. The same representative weather year and transferred demand curve are repeated over the 25-year project lifetime. PV output and EV battery capacity degrade year by year. The dispatch structure builds on previous SolarEV modelling studies and is adapted here to the global controlled-simulation setting.[3,4,39]

Hourly PV generation is simulated using the System Advisor Model, or SAM. For each grid cell, SAM generates the first-year hourly output of a standard 10 kW PV system under local meteorological conditions. The PV tilt angle is set according to the centroid latitude of the grid cell and capped at $30°$, and this cap represents a practical rooftop-deployment constraint rather than an unconstrained energy-yield optimum. PV capacity is represented by ten discrete levels, $h = 1, \dots, 10$, corresponding to 10-100% of the standardized maximum rooftop PV potential:

$$C_{g,h}^{\mathrm{PV}} = \frac{h}{10} \frac{A_g^{\mathrm{PV}}}{7.0},$$

where $A_g^{\mathrm{PV}}$ is the standardized rooftop PV potential of grid cell $g$, and $7.0\ \mathrm{m}^2\ \mathrm{kW}^{-1}$ is the area-to-capacity conversion factor. The hourly generation profile from the standard 10 kW SAM system is scaled to each candidate PV capacity and adjusted for annual degradation:

$$G_{g,h,t,y} = G_{g,t}^{\mathrm{SAM,10kW}} \frac{C_{g,h}^{\mathrm{PV}}}{10} (1 - \delta_{\mathrm{PV}})^{y-1},$$

where $y = 1, \dots, 25$, and $\delta_{\mathrm{PV}} = 0.005$. In the fixed-capacity physical simulation, all grid cells use $h = 10$.

We evaluate two operating scenarios. The PV-only scenario represents a distributed PV direct-consumption system without EV storage. PV generation first serves building electricity demand within the same hour. Surplus PV is curtailed in the main analysis. Remaining demand is supplied by the grid. The PV-EV scenario adds EV travel electricity demand and allows bidirectional vehicle-to-home operation when vehicles are connected.

In the PV-EV scenario, the nominal battery capacity of each EV is set to 40 kWh. The maximum bidirectional charge and discharge power is set to 6 kW. The aggregated EV battery capacity and maximum V2H (V2B) power in each grid cell are obtained by scaling these values linearly by the number of EVs in that grid cell. The initial battery SOC is set to 50% of nominal capacity. SOC is constrained between 50% and 95%. Round-trip efficiency is 85%. Battery aging is represented by an annual capacity-retention factor of 0.982. When retention reaches 80% or lower, the degradation counter resets and nominal capacity is restored at the next hourly update; the stored-energy state is not rescaled at replacement.

EV travel is represented by a reproducible random travel profile. For each grid cell, the aggregated EV resource was conservatively assumed to be unavailable for V2H/V2B operation during three one-hour daytime windows sampled between 07:00 and 18:00. These periods represent system-level mobility-related unavailability rather than simultaneous travel by all individual vehicles. During travel periods, the EV fleet is disconnected from the home energy system and consumes 1.1 kWh per vehicle per hour. When vehicles are connected, EV batteries can absorb surplus PV electricity. They can also discharge to serve building demand when PV generation is insufficient.

Dispatch follows a greedy rule. PV generation first serves building demand. If EVs are connected and surplus PV exists, the surplus charges the EV batteries subject to power and SOC upper-bound constraints. If EVs are connected and PV generation is insufficient, the EV batteries discharge to meet the building demand gap subject to power, efficiency, and minimum SOC constraints. At the end of each day, if EV SOC is below the 50% reserve level, the grid charges the batteries back to the travel reserve. This forced grid charging is recorded separately and is not counted as local PV self-supply. The full hourly dispatch algorithm is provided in Supplementary Information 1.5-1.7.

Self-sufficiency is evaluated from the final energy balance of the 25-year simulation. For the PV-EV configuration, the total end-use electricity demand of grid cell g is:

$$E_{\text{demand},g} = 25\left(\sum_{t=1}^{8760} D_{g,t} + 1.1 \times 3 \times 365 \times N_g^{\text{EV}}\right)$$

where $D_{g,t}$ is the transferred hourly building electricity demand and $N_g^{\text{EV}}$ is the number of EVs in grid cell g; the second term is the EV driving electricity demand, at 1.1 kWh per vehicle per travel hour, three travel hours per day and 365 days per year. The electricity actually imported from the grid over the same period is:

$$E_{\text{grid},g} = \sum_{y=1}^{25} \sum_{t=1}^{8760} I_{g,t,y}$$

where $I_{g,t,y}$ is the positive quantity of electricity actually imported from the grid in hour t of project year y. Self-supplied electricity is the difference between the two:

$$E_{\text{self},g} = E_{\text{demand},g} - E_{\text{grid},g}$$

The self-sufficiency ratio and the self-consumption ratio are then:

$$\text{SSR}_g = 100 \times \frac{E_{\text{self},g}}{E_{\text{demand},g}}$$

$$\mathrm{SC}_g = 100 \times \frac{E_{\mathrm{self},g}}{E_{\mathrm{PV},g}}$$

where $E_{\mathrm{PV},g}$ is the total PV generation of grid cell g over the 25-year period. The PV-only configuration uses the same definitions, with a denominator that contains building electricity demand only:

$$E_{\mathrm{demand},g}^{\mathrm{PV-only}} = 25 \sum_{t=1}^{8760} D_{g,t}$$

Self-supplied electricity was calculated as total end-use electricity demand minus actual electricity imported from the grid over the 25-year simulation. Total end-use demand includes building electricity demand and EV driving electricity demand in the PV-EV configuration. Grid imports include all electricity supplied by the grid, including end-of-day electricity used to restore the EV reserve state of charge. Battery round-trip efficiency remains part of the hourly dispatch calculation but is not applied again in the final self-sufficiency accounting.

**Data availability**

The meteorological inputs are the NASA GES DISC MERRA-2 hourly reanalysis products, which are publicly available from the NASA Goddard Earth Sciences Data and Information Services Center. The 2024 hourly regional load data for the Kansai Electric Power service area are published by the Organization for Cross-regional Coordination of Transmission Operators, Japan. Annual municipal electricity-demand statistics are published by the Agency for Natural Resources and Energy, Japan. Building-footprint data used to derive rooftop PV potential are published by the Geospatial Information Authority of Japan. The 2024 hourly load records for the 11 external electricity systems are obtained from the respective system operators, and the individual data sources are provided in Supplementary References 8-14. The simulated outputs that support the findings of this study, including the grid-level self-sufficiency, self-consumption and energy-sufficiency results for all 5,000 global land grid cells and all sensitivity cases, will be deposited in the Zenodo repository, and the digital object identifier will be provided at publication. These outputs are available from the corresponding authors on request during peer review. The research data will be available upon request.

**Code availability**

Hourly PV generation was simulated with the NREL System Advisor Model, which is publicly available from the National Renewable Energy Laboratory. The climate-demand transfer model and the hourly PV-EV/V2H dispatch and self-sufficiency accounting code used in this study are openly available under the MIT License at: https://github.com/liuyuchenr2-lab/PV-EV-climate-matching and are archived at Zenodo (https://doi.org/10.5281/zenodo.22771579).

**Author contributions**

T.K. conceived the study and designed the overall research framework. Y.L. developed and implemented the controlled simulation framework, performed all data processing, modelling and analysis, prepared all figures and tables, and wrote the manuscript. S.P.-L. and T.N. contributed to the early development of the study and preparation of the initial manuscript. T.K. supervised the research, contributed to the interpretation of the results, and critically revised the manuscript. All authors discussed the results and reviewed and approved the final manuscript.

**Funding**

This work was supported by JSPS KAKENHI Grant Number JP23K11520.

**Competing interests**

The authors declare no competing interests.